\documentclass[lnicst]{svmultln}
\usepackage{makeidx}  
\usepackage{graphicx}
\begin{document}
\mainmatter              
\title{Emotionally Colorful Reflexive Games}
\titlerunning{Emotionally Colorful Reflexive Games} 
%
\author{Sergey Tarasenko}
\authorrunning{Sergey Tarasenko}   
\tocauthor{Sergey Tarasenko}

\institute{Deparment of Intelligence Science and Technology, Graduate School of Informatics Kyoto University, Yoshida honmachi, Kyoto 606-8501, Japan \\
\email{infra.core@gmail.com}}

\maketitle              

\begin{abstract}        
This study addresses the matter of reflexive control of the emotional states by means of Reflexive Game Theory (RGT). It is shown how to build a bridge between RGT and emotions. For this purpose the Pleasure-Arousal-Dominance (PAD) model is adopted. The major advantages of RGT are its ability to predict human behavior and unfold the entire spectra of reflexion in the human mind. On the other hand, PAD provides ultimate approach to model emotions. It is illustrated that emotions are reflexive processes and, consequently, RGT fused with PAD model is natural solution to model emotional interactions between people. The fusion of RGT and PAD, called Emotional Reflexive Games (ERG), inherits the key features of both components. Using ERG, we show how reflexive control can be successfully applied to model human emotional states. Up to date, EGR is a unique methodology capable of modeling human reflexive processes and emotional aspects simultaneously.
\keywords{Reflexive Game Theory, Bipolar Choices, Theory of Bipolarity and Reflexivity, the Golden Ratio, Implicit Primordial Knowledge (IPK), Inner Computer, Algebraic Model of Ethical Cognition}
\end{abstract}

\section{Introduction}
\label{intro}

The theory of personal constructs proposed by Kelly \cite{kelly} in 1955 initiated the strong debates. This theory suggests that each person has one's own unique system of dichotomous (bipolar) constructs, which serve as special axes for ``projecting" self and other persons. Most of the constructs may be mapped onto the scale of ``good-bad". Kelly \cite{kelly} developed the experimental approach for essembeling the list of personal constructs and a list of one's images of other people. The poles, the constructs are represented with, are of negative (``bad") and positive (``good") types. Hereafter, the theory proposed by Kelly will be referred as \textit{Theory of Bipolar Constructs}. According to Kelly, these poles should be chosen with equal probability. However the subsequent psychological experiments proved this suggestion wrong.

In earlier experiments, it was shown by Benjafield and Adams-Webber\cite{adamsben2}, Adams-Webber\cite{adams1}, Benjafield and Green \cite{bengreen},  Shalit \cite{shalit}, Osgood and Richards \cite{osgoodrich} and Osgood \cite{osgood}  that under great variety of experimental conditions the positive pole is chosen with the frequency 0.62.

The persistent appearance of the frequency 0.62 resulted in two ways. First, the reasonable question "What is the true value of the probability this frequency represents?". Second, it was suggested by Lefebvre \cite{lef3}(p. 291) that 
\begin{quotation}
``... human cognition has a special mechanism for modeling self and others. It works as a universal `inner computer' and creates the core of the images that later are `dressed' and  `colored' with nuances. The constant 0.62 is a characteritic of this `computer'."
\end{quotation}

It was determined by Lefebvre that the true value of the frequency 0.62 is probability 0.618..., which is inverted Golden Ratio $\phi$ ($\phi$ = 1.618...) \cite{lef3,lef5}.

Therefore, the Golden Ratio, as being a characteristic of the `inner computer', is related to the Implicit Primordial Knowledge (IPK) \cite{taraset}.

Now we return to the Theory of Bipolar Constructs. It had strong impact on subsequently conducted psychological researches. The research about the emotions was not an exception. 

The semantic differential approach originally proposed by Osgood et al. \cite{osgoodet} considers three dimensions to characterize the person's personality. These dimensions are Evalution, Activity and Potency. 

This approach was further tested from the point of view of emotions. Russel and Mehrabian \cite{rusmeh} proved in their study that the entire spectra of emotions can be described by the 3-dimensional space spanned by Pleasure (Evaluation), Arousal (Activity) and Domination (Potency) axes. For every dimension, the lower and upper bounds (ends) are recognized as negative and positive poles, respectively. Consequently, the negative pole can be described by negative adjectives, and positive one by positive adjectives. 

It was shown by Russel and Mehrabian \cite{rusmeh} that these three dimensions are not only necessary dimensions for an adequate description of emotions, but they are also sufficient to define all the various emotional states.  In other words, the \textit{Emotional state} can be considered as a function of $Pleasure$, $Arousal$ and $Dominance$. On the basis of this study, Mehrabian \cite{meh} proposed Pleasure-Arousal-Dominance (PAD) model of Emotional Scales.  The emotional states defined as combinations of ends from various dimensions are presented in Fig.~\ref{pad}. 

In this study, we discuss the matter of how the PAD model can be used in Reflexive Game Theory (RGT) \cite{lef1,lef2} to emotionally color the interactions between people and humans and robots.

\section{Building the Bridge between PAD Model and RGT}

\subsection{Representation of PAD model in the space of binary 3D vectors}
The most important issue to fuse RGT inference and PAD Emotional Scale is to build the bridge between two approaches. 

\begin{figure}
\centering
\includegraphics[width=11.5cm]{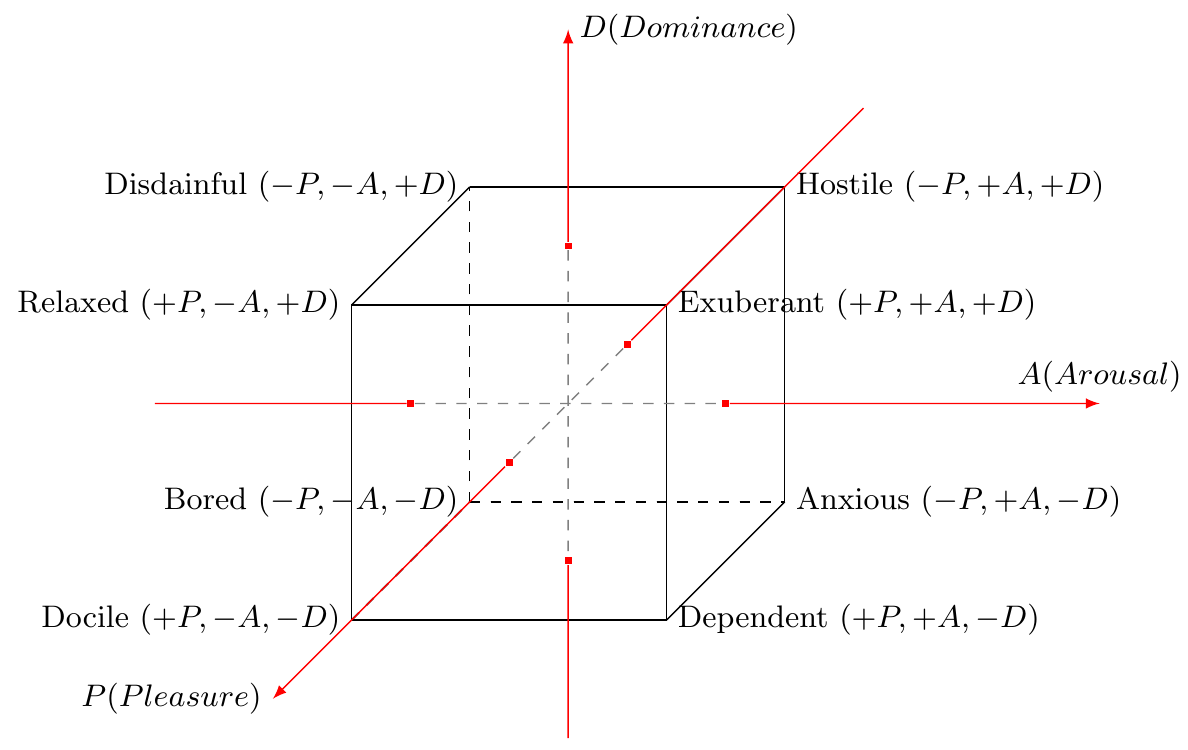}
\caption{The Pleasure-Arousal-Dominance (PAD) model's space.}
\label{pad}
\end{figure}

By definition, PAD model is spanned by three dimensions. The value of each component continuously ranges from -1 to 1. The notation in the PAD model space presented in Fig.~\ref{pad} are as follows: 1) pair $+P$ vs $-P$ corresponds to $Pleasure$  (positive pole: value 1) vs $Displeasure$ (negative pole: value -1); 2) pair $+A$ vs $-A$ corresponds to $Arousal$ (positive pole: value 1) vs $Non$-$arousal$ (negative pole: value -1); and 3) pair $+D$ vs $-D$ corresponds to $Dominance$ (positive pole: value 1) vs $Submissiveness$ (negative pole: value -1).

 Accodring to Mehrabian \cite{meh}, ``pleasure vs displeasure" distinguishes the positive vs negative emotional states, ``arousal vs non-arousal" refers to combination of physical activity and mental alertness, and ``dominance vs submissiveness" is defined in terms of control vs lack of control \cite{meh}.

Though, PAD model operates with continous values, i.e., for example, $curious$ is coded as $(0.22, 0.62, -0.01)$. Mehrabian \cite{meh} defines 8 basic states, which are all possible combinations of high vs low pleasure ($+P$ vs $-P$), high vs low arousal  ($+A$ vs $-A$), and high vs low dominance ($+D$ vs $-D$). In other words, there 8 basic states are all possiblecombinations of the poles. In total, there are 6 poles - two for each of three scales. Therefore, there are $2^3 = 8$ possible combinations of poles. These states (combinations) are considered as extreme states and can be refered as approximations of the intermediate states. For instance, intermediate state ``angry" (-0.51, 0.59, 0.25) can be approximated by extreme state ``hostile" $(-P,+A,+D)$. Therefore, only eight extreme states can be used to describe the entire variety of intermediate emotional states.

\begin{table}
\caption{Disjunction operation}
\begin{center}
\begin{tabular}{|c|c|c|}
\hline
x&y& x $\cup$ y \\
\hline
\rule{0pt}{12pt}0&0&0\\[2pt]
\hline
\rule{0pt}{12pt}1&0&1\\[2pt]
\hline
\rule{0pt}{12pt}0&1&1\\[2pt]
\hline
\rule{0pt}{12pt}1&1&1\\[2pt]
\hline
\end{tabular}
\end{center}
\label{dis}
\end{table}

\begin{table}
\caption{Conjunction operation}
\begin{center}
\begin{tabular}{|c|c|c|}
\hline
x&y& x $\cap$ y \\
\hline
\rule{0pt}{12pt}0&0&0\\[2pt]
\hline
\rule{0pt}{12pt}1&0&0\\[2pt]
\hline
\rule{0pt}{12pt}0&1&0\\[2pt]
\hline
\rule{0pt}{12pt}1&1&1\\[2pt]
\hline
\end{tabular}
\end{center}
\label{con}
\end{table}

The combinations of poles in the PAD space can be presented as 3-dimensional binary vectors, where 1 corresponds to positive pole (1 in PAD notation) and 0 refers to negative pole (-1 in PAD notation). Therefore, emotional states, which are combinations of different poles of various dimensions, can be represented as follows: for instance, emotional state \textit{Hostile (-P,+A,+D) or (-1,1,1)} can be substituted for $\{0,1,1\}$, while \textit{Relaxed (+P,-A,+D) or (1,-1,1)} can be represented as $\{1,0,1\}$ in the rescaled coordinates.

The complete set of emotional states represented as binary vectors is \\
$Docile$ $(+P,-A,-D)$ is coded as $\{1,0,0\}$;\\
$Anxious$ $(-P,+A,-D)$ is coded as $\{0,1,0\}$; \\
$Disdainful$ $(-P,-A,+D)$ is coded as $\{0,0,1\}$; \\
$Hostile$ $(-P,+A,+D)$ is coded as $\{0,1,1\}$;\\
$Dependent$ $(+P,+A,-D)$ is coded as \{1,1,0\};\\
$Relaxed$ $(+P,-A,+D)$ is coded as $\{1,0,1\}$;\\
$Exuberant$ $(+P,+A,+D)$ is coded as $\{1,1,1\}$;\\
$Bored$ $(-P,-A,-D)$ is coded as $\{0,0,0\}$.

Among eight basic states, there are three special emotional states  \textit{Docile (+P,-A,-D)} $\equiv \{1,0,0\}$), $Anxious$ $(-P,+A,-D) \equiv \{0,1,0\}$)  and $Disdainful$ $(-P,-A,+D) \equiv \{0,0,1\}$). These three states are the basis of the 3d binary space. Thus, any other five emotional states can be considered as disjunction (notations OR/$\cup$/+) of these three basis vectors (emotional states). The binary disjunction operation is presented in Table \ref{dis}.

For example, emotional state $Dependent$ $(+P,+A,-D)$ is disjunction of basis states $Docile$ $(+P,-A,-D)$ and $Anxious$ $(-P,+A,-D)$: $Docile$ OR $Anxious$ = $(+P,-A,-D)$ $\cup$ $(-P,+A,-D)$ = $\{1,0,0\} \cup \{0,1,0\}$ = $\{1,1,0\}$ = $(+P,+A,-D)$ = Dependent.

\subsection{Formalism of the Reflexive Game Theory}
In this section, we present the basics of the  Reflexive Game Theory (RGT). In the RGT, it is considered that behavior of each subject in the group is determined by \textit{reflexive function} $\Phi$, which is function of subjects' mutual influences \cite{lef1,lef2,lef4,lef5}:
\begin{equation}
\Phi = \Phi(a_1,...,a_i,...,a_n), 
\label{reffun}
\end{equation}
where $n$ is a total number of subjects in the group; $a_j$, for any $j=1,...,n; \ j \neq i$, influence of subject $a_j$ on subject $a_i$.

Then the choice of subject $a_i$ is defined by \textit{decision equation} \cite{lef1,lef2}:
\begin{equation}
a_i = \Phi(a_1,...,a_i,...,a_n), 
\label{reffun}
\end{equation}

The RGT uses Boolean algebra of alternatives for calculus. There three operations defined in Boolean algebra. These are unary negation operation and binary conjunction and disjunction operations. The disjuction and conjunction operations are defined in Tables \ref{dis} and \ref{con}, respectively.

In this study, we employ Boolean algebra containing eight 3d binary vectors: $\{1,0,0\}$, $\{0,1,0\}$, $\{0,0,1\}$, $\{1,1,0\}$, $\{1,0,1\}$, $\{0,1,1\}$, $\{0,0,0\}$, and $\{1,1,1\}$. The element of Boolean algebra code particular influences. Each element of Boolean algebra is referred as $alternative$.

The RGT deals with groups of individuals. Any group is represented with \textit{relationship graph}. Any group, whose graph is decomposable\footnote{See \cite{lef1,lef2,lef4} for detailed discussion about graph decomposition.}, can be represented as polynomial $P$. 

The relationships in the group are indicated with ribs. The solid-line ribs represent alliance relationship. The dashed-line ribs correspond to conflict relationship. In algebraic form, conjunction or multiplication operations represent alliance, while disjunction or summation operations correspond to conflict relationship. The idea behind is as follows. Two subjects being in alliance can find the compromise or a common influence, therefore their interaction can be characterized as conjunction of their influences. At the same time, two conflicting subjects will never reach a compromise and their interaction should include both influences. Therefore disjunction operation corresponds to the conflict relationship. 

\begin{figure}
\centering
\includegraphics[width=2cm]{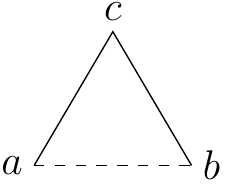}
\caption{Relationship graph for three subject $a$, $b$ and $c$.}
\label{relgr}
\end{figure}

Considerring the disjunction and conjunction operations, the group, in which subjects $a$ and $b$ are in conflict with each other and in alliance with subject $c$, is represented as $(a+b)c$ polynomial. Relationship graph is presented in Fig.~\ref{relgr}.

\begin{figure}
\centering
\includegraphics[width=2cm]{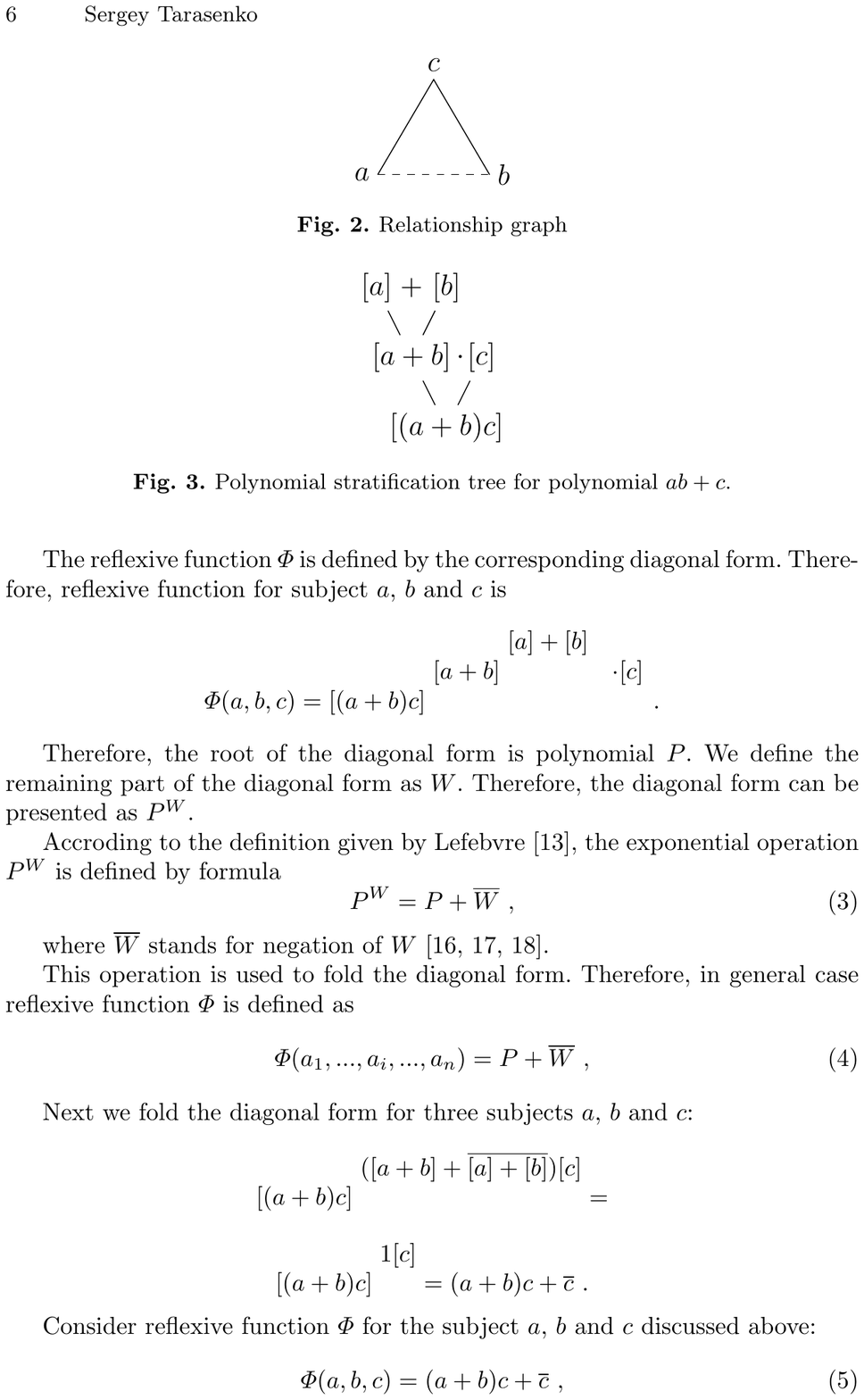}
\caption{Polynomial stratification tree for polynomial $(a+b)c$.}
\label{pst}
\end{figure}

Then on the basis of polynomial $P$, the \textit{polynomial stratification tree} is constructed (Fig.~\ref{pst}).

By omiting the branches of the polynomial stratification tree, the \textit{diagonal form} \cite{lef1,lef2,lef5,lef6,taras1} is obtained straight forwardly. Consider the diagonal form corresponding to the PST presented in Fig.~\ref{pst}:
\[\begin{array}{*{20}{c}}
   {} & {} & {[a]+[b]} & {}  \\
   {} & {[a+b]} & {} & {\cdot[c]}  \\
   {[(a+b)c]} & {} & {} & {\ \ \ \ \ \ \ .}  \\
\end{array}\]

The reflexive function $\Phi$ is defined by the corresponding diagonal form. Thus, reflexive function for subjects $a$, $b$ and $c$ is 

\[\begin{array}{*{20}{c}}
   {} & {} & {[a]+[b]} & {}  \\
   {} & {[a+b]} & {} & {\cdot[c]}  \\
   {\Phi (a,b,c) = [(a+b) c]} & {} & {} & {\ \ \ \ \ \ \ .}  \\
\end{array}\]

Therefore, the root of the diagonal form is polynomial $P$. We define the remaining part of the diagonal form as $W$. Thus, the diagonal form can be presented as $P^W$.

Accroding to the definition given by Lefebvre \cite{lef5}, the exponential operation  $P^W$ is defined by formula 
\begin{equation}
\label{expfrom}
P^W = P + \overline{W} \ , 
\end{equation}
where $\overline{W}$  stands for negation of $W$ \cite{lef1,lef2,lef4}. 

This operation is used to fold the diagonal form. Therefore, in general case reflexive function $\Phi$ is defined as 
\begin{equation}
\Phi(a_1,...,a_i,...,a_n)  = P + \overline{W} \ , 
\label{reffunr}
\end{equation}

Next we fold the diagonal form for three subjects $a$, $b$ and $c$:
\[\begin{array}{*{20}{c}}
   {} & {([a+b] + \overline{[a]+[b]}) [c]} & {}   \\
   {[(a+b)c]} & {} & {=}  \\
\end{array}\]
\[\begin{array}{*{20}{c}}
   {} & {1[c]} & {}   \\
   {[(a+b)c]} & {} & {=(a+b)c + \overline{c} \ .}  \\
\end{array}\]

The reflexive function $\Phi$ for the subjects $a$, $b$ and $c$ discussed above is then:
\begin{equation}
\Phi(a,b,c)  = (a+b)c + \overline{c} \ , 
\label{reffunr1}
\end{equation}

The general form of decision equation for each subject is then
\begin{equation}
x  = (a+b)c + \overline{c} \ , 
\label{reffunr2}
\end{equation}
where $x$ is any subject variable $a$, $b$ or $c$.

Finally, we briefly present the solution method of the decision equation. Consider the decision equaition in a form:
\begin{equation}
x  = Ax + B\overline{x}\ , 
\label{reffunr2}
\end{equation}
where $A$ and $B$ are some sets.

This form of the decision equation is called \textit{canonical form} \cite{lef1,lef2,taras1,taras2} of the decision equation. If sets $A$ and $B$ are such that $A \supseteq B$, then decision equation in canonical form has at least one solution from the interval $A \supseteq x \supseteq B$.

\subsection{Merging the RGT and PAD model}
Summarizing the facts about the RGT and PAD model, we highlight that RGT has been proven to predict human choices in the groups of people and allows to control human behavior by means of particular influences on the target individuals. Next we note that PAD model provides description of how the emotional states of humans can be modelled, meaning that a certain emotional state of a particular person can be changed to the desired one. Furthermore, it is straightforward to see that the coding of the PAD emotional states and alternatives of Boolean algebra are identical.

Therefore, it is possible to change the emotional states of the subjects in the groups by making influences as elements of the Boolean algebra. In such a case, vector $\{1,0,0\}$, for example, plays a role of influence towards emotional state Docile.

Besides, we have distinguished three basis emotional states $Docile$ ($\{1,0,0\}$), $Anxious$ ($\{0,1,0\}$) and $Disdainful$ ($\{0,0,1\}$). The interactions (as defined by disjunction and conjunction operations) of these basic emotional states can result in derivative emotional states such as $Dependent$, $Relaxed$, etc. Before, considering the example of PAD application in RGT, we note that reflexive function $\Phi$ defines state, which subject is going to switch to. This process goes unconsciously. We have discussed above the reasons the conjunction and disjunction represent alliance and conflict relatioships, respectively. 

Another important issue is that often people directly express their emotions in actions. Therefore in such a case, a particular emotional state of a subject can be considered as the influence he is making on other subjects.

\textit{Example 1}. Subjects $a$ and $b$ are in alliance relationship. Subject $a$ makes influence $Dependent \ \{1,1,0\}$. Subject $b$ makes influence $Relaxed \ \{1,0,1\}$. Their resultant influence will be $(a \cdot b) = \{1,1,0\}\{1,0,1\} = \{1,0,0\}$ or $Docile$. Consequently, the influence of the group, including subjects in alliance with each other, on a given subject is considered as conjunction (defining compromise of all the subjects in alliance) of the influences of all the subjects' influences.

\textit{Example 2}. Subjects $a$ and $b$ are in conflict relationship. Subject $a$ makes influence $Docile \ \{1,0,0\}$. Subject $b$ makes influence $Disdainful \ \{0,0,1\}$. Their resultant influence will be $(a + b) = \{1,0,0\}\{0,0,1\} = \{1,0,1\}$ or $Relaxed$. Therefore, the influence of the group, including subjects in conflict with each other, on a given subject is considered as disjunction (defining overall influence since compromise is impossible) of the influences of all the subjects' influences.

Next, we consider an example of reflexive interactions controlling emotional states.

\section{Emotionally Colored Reflexive Games: Sample Situation}
\begin{figure}
\centering
\includegraphics[width=2cm]{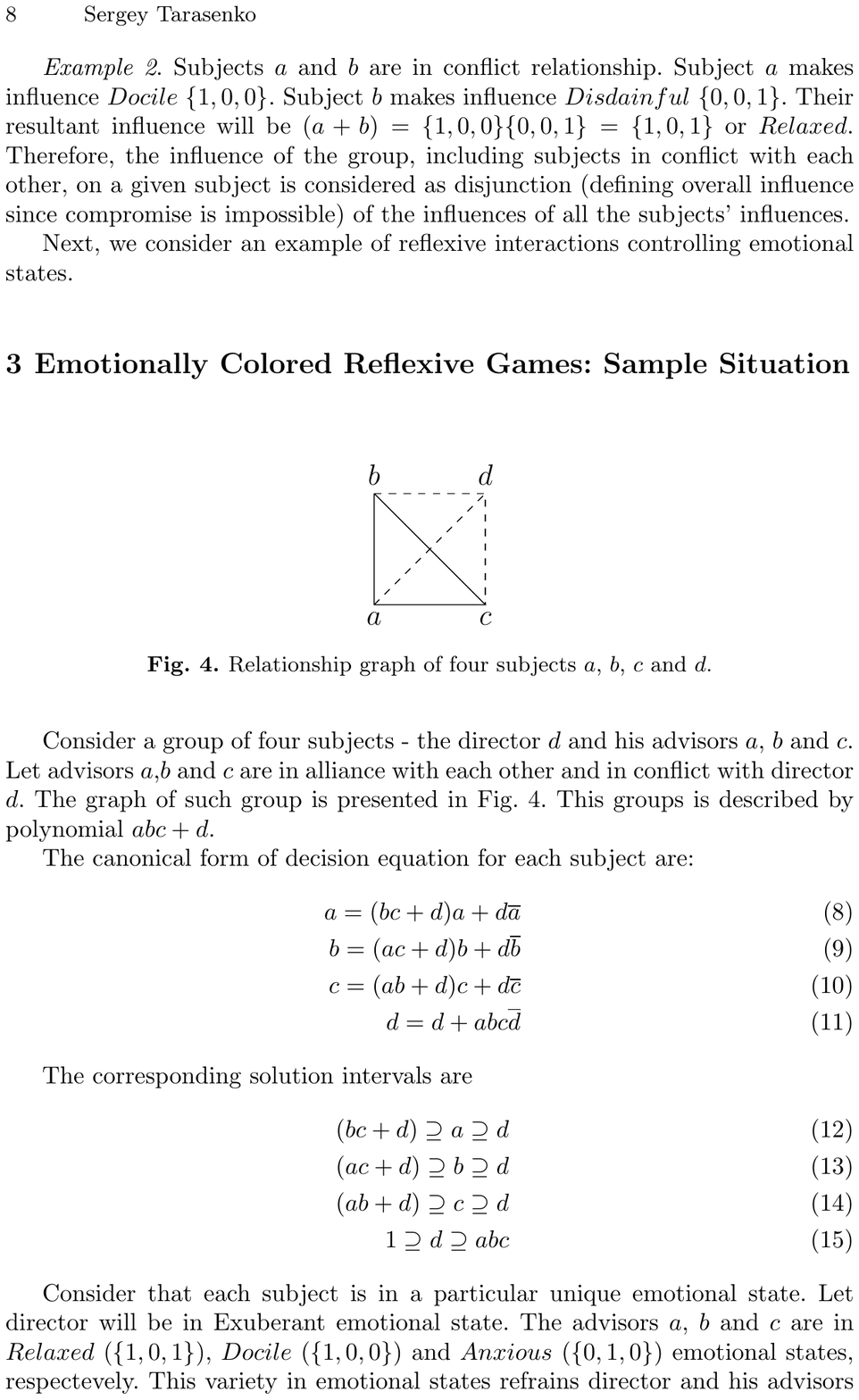}
\caption{Relationship graph of four subjects $a$, $b$, $c$ and $d$.}
\label{foursub}
\end{figure}

Consider a group of four subjects - the director $d$ and his advisors $a$, $b$ and $c$. Let advisors $a$,$b$ and $c$ are in alliance with each other and in conflict with director $d$. The graph of such group is presented in Fig.~\ref{foursub}. This groups is described by polynomial $abc+d$.

The canonical form of decision equation for each subject are:
\begin{eqnarray}
a = (bc+d)a + d\overline{a} \\
b = (ac+d)b + d\overline{b} \\
c = (ab+d)c + d\overline{c} \\
d = d + abc\overline{d}
\end{eqnarray}

The corresponding solution intervals are
\begin{eqnarray}
(bc+d) \supseteq a \supseteq d \\
(ac+d) \supseteq b \supseteq d \\
(ab+d) \supseteq c \supseteq d \\
1 \supseteq d \supseteq abc
\end{eqnarray}

It is assumed that each subject is in a particular unique emotional state. Let director will be in Exuberant emotional state. The advisors $a$, $b$ and $c$ are in $Relaxed$ ($\{1,0,1\}$), $Docile$ ($\{1,0,0\}$) and $Anxious$ ($\{0,1,0\}$) emotional states, respectevely. This variety in emotional states refrains director and his advisors from reaching a fruitful decision. Understanding this emotional situation, director decides to apply reflexive control on emotional level. Let adivors' influence on all the other subjects coincides with their emotional state, while directly is in complete control and can decide, which emotional influence to on each particular subject. 

Using RGT, we can predict the emotional states of each subject in the group after the \textit{reflexive emotional interaction}:\\
for subject $a$: $(\{1,0,0\}\{0,1,0\}+d) \supseteq a \supseteq d \Rightarrow a = d$; \\
for subject $b$: $(\{1,0,1\}\{0,1,0\}+d) \supseteq b \supseteq d \Rightarrow a = d$; \\
for subject $c$: $(\{1,0,1\}\{1,0,0\}+d) \supseteq c \supseteq \Rightarrow \{1,0,0\}+d \supseteq c \supseteq d $; \\
for subject $d$: $1 \supseteq d \supseteq \{1,0,1\}\{1,0,0\}\{0,1,0\} \Rightarrow 1 \supseteq d \supseteq 0 \Rightarrow d = d$.

Therefore, under conditions of such group structure and influences, decision of advisors $a$ and $b$ is completely defined by the director's influence. 
\begin{figure}
\centering
\includegraphics[width=7cm]{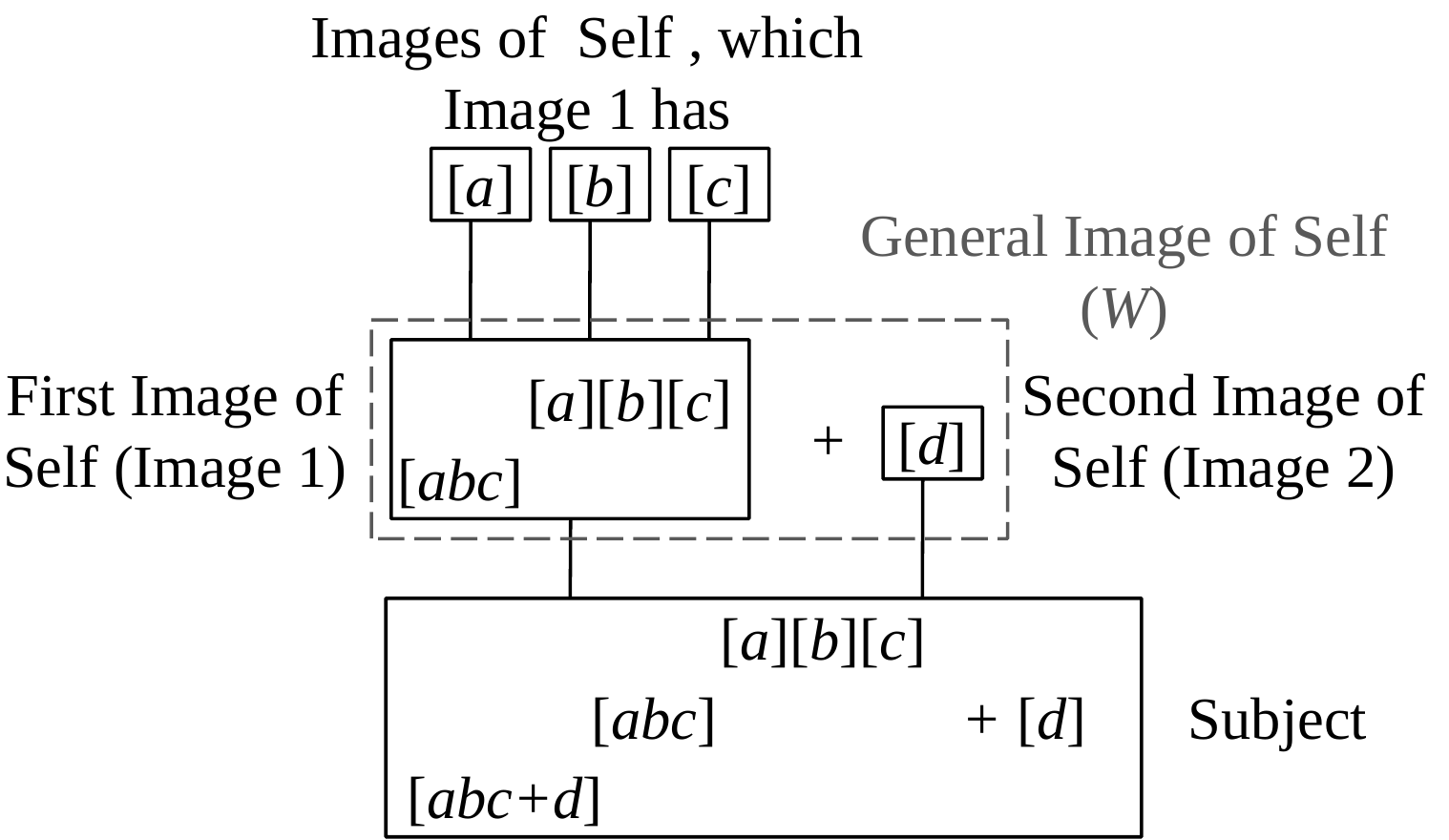}
\caption{Interpretation of the Diagonal form levels.}
\label{images}
\end{figure}

The entire diagonal form ($P+\overline{W}$) represents the state of the subject. In Fig.~\ref{images}, the diagonal form is marked as Subject.  The term $\overline{W}$ is called a general image of the self. On the next level, there are two images of Image 1 ($[abc]^{[a][b][c]}$) and Images 2 ($[d]$). The Images 1 and 2 are images of the self, which general image of the self $W$ has. Finally, there are the images $[a]$, $[b]$ and $[c]$ are the images of the self, which Image 1 has.

Following this interpretation of the diagonal form \cite{lef1,lef2}, we can calculate each emotional state in each image. We analyze the structure of reflexion for advisor $c$. His state is ($\{1,0,0\} + d$) or ($Docile(+P,-A,-D)$ plus director's influence).  The emotional state in the Image 1 is $Exuberant$ ($1 = \{1,1,1\}$), because $[abc]^{[abc]} = [abc] + \overline{[abc]} = 1$. The Image 2 is $[d]$ that means it is entirely defined by director's influence. Finally, the general images of the self $W$ is $Exuberant$ ($1 = \{1,1,1\}$). 

In this simple example, we have illustated how the emotional states can appear to be the subject for the reflexive control. We call such reflexive control to be \textit{reflexive emotional control}. We have shown how the reflexive emotional control can be sucessfully implemented by means of the Reflexive Game Theory.

Besides, the RGT allows to unfold the entire sequence of reflexion in the human mind including its emotional aspects. 

\section{Discussion}
In the previous sections, we have introduced the bridge between the PAD model of emotional states and RGT calculus by coding PAD's emotoinal states in terms of RGT's Boolean algebra of alternatives. We also illustrated how the emotional states can be the subject of reflexive control and can be successfully managed. This is possible, because the variables in the PAD, namely, Pleasure, Arousal, Dominance, are proved to be necessary and sufficient variables to control human emotions \cite{rusmeh}. In other words, necessity means that if emotional state is changed the value of all three variables is changed accordingly. On the other hand, if the values of the variables are changed, the emotional state is also changed. We have illustrated how to merge the RGT calculus with proper coding of PAD emotional states. Finally, we have provided a simple explanatory example of how reflexive emotional control can be applied in action.

Furthermore, we have not only illustated how to apply RGT to the control of subject's emotions, but uncovered the entire cascade of human reflexion as a sequence of \textit{subconscious reflexion}, which allows to trace emotional reflection of each reflexive image. This provides us with unique ability to unfold the sophisticated structure of reflexive decision making  process, involving emotions. Up to date, there has been no approach, capable of doing such thing, reported. 

The emotional research based on PAD model is transparent and clear. The models of robots exhibiting human like emotional behavior using only PAD has been successfully illustrated in recent book by Nishida et al. \cite{nishida}. The ability to influence the factors (variables) is the major justification for such approaches and for the application of the RGT. Yet, it is just a ``mechanical'' part of the highly sophisticated matter of human emotions. 

The present study introduces the brand new approach to modeling of human emotional behavior. We call this new breed of RGT application to be \textit{emotional Reflexive Games (eRG)}. At present RGT fused with PAD model is the unique approach allowing to explore the entire diversity of human emotional reflexion and model reflexive interaction taming emotions. This fusion is automatically formalized in algorithms and can be easily applied for further developing emotional robots.  

Since the proposed mechanism has no heavy negative impact on human psychologycal state, robots should be enabled to deal with such approach in order to provide human subjects with stress free psychological friendly environments for decision making.

Next, we discuss more complex question regarding emotion and reveal the facts of why the Reflexive Game Theory is the natural approach to model emotions.

We start from trying to answer the question - what are the emotions themselves? 

According to Ekman \cite{ekman2}, emotions shoud be regarded as highly automatics processing algorithms responsible for our everyday life survival. Being entire unconscious (automatic) regulators of the human body's physiology, emotions bring the instantenous solutions of how to act in front of rapidly approaching possible threat. This instantenous activity is possible due to unconscious processing algorithms, which have been characterized by Lewicki et al. \cite{lewicki1,lewicki2} as highly non-linear by nature and extremely fast by performance.

Therefore, emotions are fast processors of the information coming apart from environment. Usually, the emotions are characterized by some physiological patterns of body activity on the one hand, and external expression by face mimic or gestures on the other hand. From this point of view, the fact that the reproduction by the self of physical part of emotional pattern, i.e., just making an angry face can elicit the anger as emotional state itself \cite{ekman,leven}, is completely scientifically unexpected phenomenon. Yet this seems to be widely known by writers. For instance, Edgar Allan Poe, in his story ``The Purloined Letter'', describes how one character is attemping to understand the intensity of emotional experience of another character by self-mimicing (or imitating) facial expressions of another one. 

This imitation is interesting from the point of view of reflexive processes. Vladimir Lefebvre in his book ``Structure of Awareness" \cite{lef9} writes:
\begin{quotation}
``Reflexion in traditional phylosophical psychological sense is ability to place one's mind into external position of `observer', `researcher' or `controler' of one's own body, actions and thoughts. We extend such understanding of reflexion and will consider that reflexion is ability to place one's mind into external position to another `character', his actions and thoughts.".
\end{quotation}

From this point of view, the self-mimicing is reflexive process. Thus, it is possible to elicit and understand the emotions of others by $reflexion$. Ekman et al. \cite{ekman}(p. 1210) suggested that this is possible due to ``... direct connections between motor cortex and hypothalamus that traslates between emotion-prototypic expression in the face and the emoition-specific pattering in the autonomic nervous system".  

On the other hand, the reflexive processes are at the core of the RGT. Therefore RGT is natural approach to model emotions.

Continuing the line of logic that emotions can be elicited by self reproducing patterns of motor activity accompanying expression of emotions, we suggest the not only facial expressions, but also other motor patterns such as gestures can elicit the corresponding emotion.

Consequently wide variaty of motor actions can cause some emotional states to emerge. On the other hands, other actions, not related to the physical activity, can result in particular emotions. Therefore, it is possible to control human emotions not directly, but making people to perform particular actions. From this perspetive, PAD model can serve as a measuring device of how actions influence emotions.

However, since the emotions are automatic processing algorithms, which take their input apart from environment, the environment itself is natural medium to influence on emotions. How this can be applied in interior of shops to provide a customer with delightful emotional state is described by Soriano et al. \cite{soriano}.

Therefore, the continously on-going decision making process of selecting and implementing action etc. is always accomapanied by emotional process, which characterize each action from the emotional perspective.

We consider that the material presented in this study has essential influence on the brand new Reflexive Game Theory. Furthermore, the results presented extend widely recognized field of Affective Compiting \cite{picard} with new powerful computational paradigm, which is ready to be applied in robots.




\end{document}